\documentclass[graybox]{svmult}

\usepackage{mathptmx}       % selects Times Roman as basic font
\usepackage{helvet}         % selects Helvetica as sans-serif font
\usepackage{courier}        % selects Courier as typewriter font
\usepackage{type1cm}        % activate if the above 3 fonts are
\usepackage{makeidx}         % allows index generation
\usepackage{graphicx}        % standard LaTeX graphics tool
\usepackage{multicol}        % used for the two-column index
\usepackage{url}             % allows inclusion of URLs
\usepackage[bottom]{footmisc}% places footnotes at page bottom
\usepackage{float, amsmath, amssymb, natbib, pdfsync}
\usepackage{amsbsy}
\makeindex             % used for the subject index
\begin{document}
\title*{A Tale of Two Current Sheets$^*$}

\author{
Jonathan Arons
}

\institute{
Department of Astronomy, Department of Physics, Space Sciences Laboratory and Theoretical Astrophysics Center \\ University of California, Berkeley\\ 
\email{arons@berkeley.edu} \\
\newline
\hspace*{-0.03in} *With apologies to Charles Dickens
}

\maketitle

%\abstract{
%Each chapter should be preceded by an abstract (10--15 lines long) that summarizes the content. The abstract will appear \textit{online} at \url{www.SpringerLink.com} and be available with unrestricted access. This allows unregistered users to read the abstract as a teaser for the complete chapter. As a general rule the abstracts will not appear in the printed version of your book unless it is the style of your particular book or that of the series to which your book belongs.
%Please use the 'starred' version of the new Springer \texttt{abstract} command for typesetting the text of the online abstracts (cf. source file of this chapter template \texttt{abstract}) and include them with the source files of your manuscript. Use the plain \texttt{abstract} command if the abstract is also to appear in the printed version of the book.
%}

\abstract{I outline a new model of particle acceleration in the current sheet separating the closed from the open field lines in the force-free model of pulsar magnetospheres, based on reconnection at the light cylinder and ``auroral'' acceleration occurring in the return current channel that connects the light cylinder to the neutron star surface.  I discuss recent studies of Pulsar Wind Nebulae, which find that pair outflow rates in excess of those predicted by existing theories of pair creation occur, and use those results to point out that dissipation of the magnetic field in a pulsar's wind upstream of the termination shock is restored to life as a viable model for the solution of the ``$\sigma$'' problem as a consequence of the lower wind 4-velocity implied by the larger mass loading.\renewcommand{\thefootnote}{\fnsymbol{footnote}}
\footnote[2]{Collaborators, none of whom should be held responsible for the content of this paper: D. Alsop, E. Amato, D. Backer, P. Chang, N. Bucciantini, B. Gaensler, Y. Gallant, V. Kaspi, A.B. Langdon, C. Max, E. Quataert, A. Spitkovsky, M. Tavani,  A. Timokhin}} 
\renewcommand{\thefootnote}{\arabic{footnote}}

\section{Follow The Energy}\label{sec:intro}

Rotation Powered Pulsars (RPPs) provide the first and most definitive example of compact astrophysical systems which draw the power for their observed emissions from the extraction of rotational energy from  gravitationally bound objects through the action of macroscopic electromagnetic fields.  They have motivated models for similar energy extraction from disks around other gravitating bodies, such as black holes (e.g. \citet{rees84, begelman84}). As objects of study, the RPPs have a virtue lacking in the black hole systems: timing of the precisely measured pulse periods, uniquely interpretable as the rotation periods of the underlying neutron stars, provide measurements of the total energy budget free of all astrophysical uncertainties, other than the factor of $\sim 2$ uncertainty in neutron stars' moments of inertia, arising from the unecertainties in the equation of state of dense matter. The measured rate of rotational energy loss, 
\begin{equation}
\dot{E}_R = -I \Omega_* \dot{\Omega}_* = 4\pi I \frac{\dot{P}}{P^3},
\label{eq:spinloss}
\end{equation}
tells us the total energy budget for these systems, without our having to understand anything about the photon emissions from these systems - which is both a blessing and a curse - a blessing, because in contrast to other relativstic astrophysical systems, we know the energy budget, without having to unravel the partition between flow kinetic energy, large scale Poynting flux, thermal energy and radiative losses - a curse, since the energy loss is radiatively silent, thus supplying little information as to the details of the energy outflow, leaving the mechanics of the machine mysterious.  

Nevertheless, progress has been made. The Pulsar Wind Nebulae (PWNe) act as catch basins for the rotational energy lost.  Observations of these systems \citet{gaensler06}, using radio (including millimeter), X-ray, gamma ray and occasionally infrared telescopes\footnote{Most PWNe lie in the galactic plane, therefore are relatively inaccessible to optical techniques, and are even less accessible to UV telescopes.  Near and far infrared observations are extremely useful in unraveling the physics of the relativistic outflows \citep{bucc10}, but have been much less in evidence than the high energy studies.}, have made clear that RPPs deliver their energy to the outside world in the form of highly relativistic, magnetized outflows - stellar winds that are exaggerated versions of the solar wind - which must be electromagnetically driven by the magnetic pressure of the wound up magnetic field. The strength of that field is estimated by using the theory of magnetic braking of the neutron stars' spin, which suggests 
\begin{equation}
\dot{\Omega}  =  -K \Omega^n, \; \Omega = 2\pi /P, 
\label{eq:spindown}
\end{equation}
$ P = $ rotation period, applied to the observed rotations periods and spindown rates $\dot{P} = - 2\pi \dot{\Omega}/\Omega^2$. 

The earliest model applied vacuum electrodynamics to a rotating sphere endowed with a magnetic dipole moment ${\boldsymbol \mu}$ centered at the stars' centers and tipped with respect to the rotation axis by an angle $i$.  That theory yields the spindown luminosity $\dot{E}_R = K \Omega^4, \; K = (2/3)\mu^2 \sin^2 i /c^3$ [{\it e.g.} \cite{pacini67, ostriker69}] thus $n = 3$ in this model. Vacuum theory was motivated by the large gravitational forces at the surfaces, suggesting no plasma more than a meter or so above the star, but immediately Deutsch's much earlier observation [\cite{deutsch55}], made in the context of magnetic A stars, that in vacuum large electric fields parallel to $B$ would overwhelm gravity and pull charged particles out from the star until the vacuum electric field would be altered, reducing ${\boldsymbol E \cdot B}$ down to zero, was recovered \citep{gold69} and extended with the suggestion that the charged particles would feed a curious charge separated wind, with a total electric current $I = c \Phi_{\rm mag}, \; \Phi_{\rm mag} = $ total magnetospheric potential $= \sqrt{\dot{E}_R /c}$.  That wind could carry away the rotational energy of even the aligned rotator, in a Poynting flux dominated flow - the electromagnetic energy density would vastly exceed the kinetic energy density (and pressure) of the outflow, in the initially conceived model - the particle flux in that scheme is only $c\Phi/e = 2.3 \times 10^{30} (I_{45} \dot{P}_{15}/P^3)^{1/2} $ elementary charges/s, $I_{45} = I/10^{45} \; {\rm cgs}, \dot{P}_{15} =  \dot{P}/10^{-15}$, the ``Goldreich-Julian'' current. 

The charge separated model has several really serious theoretical difficulties, but perhaps of greater importance is that the observations of {\it young} PWNe have made 
%\vspace*{-1cm}
\begin{figure}[H]
\begin{center}
%\unitlength = 0.0011\textwidth
\hspace{10\unitlength}
\begin{picture}(300,100)(0,15)
\put(0,-100){\makebox(300,200)[tl]{\includegraphics[width=4in]{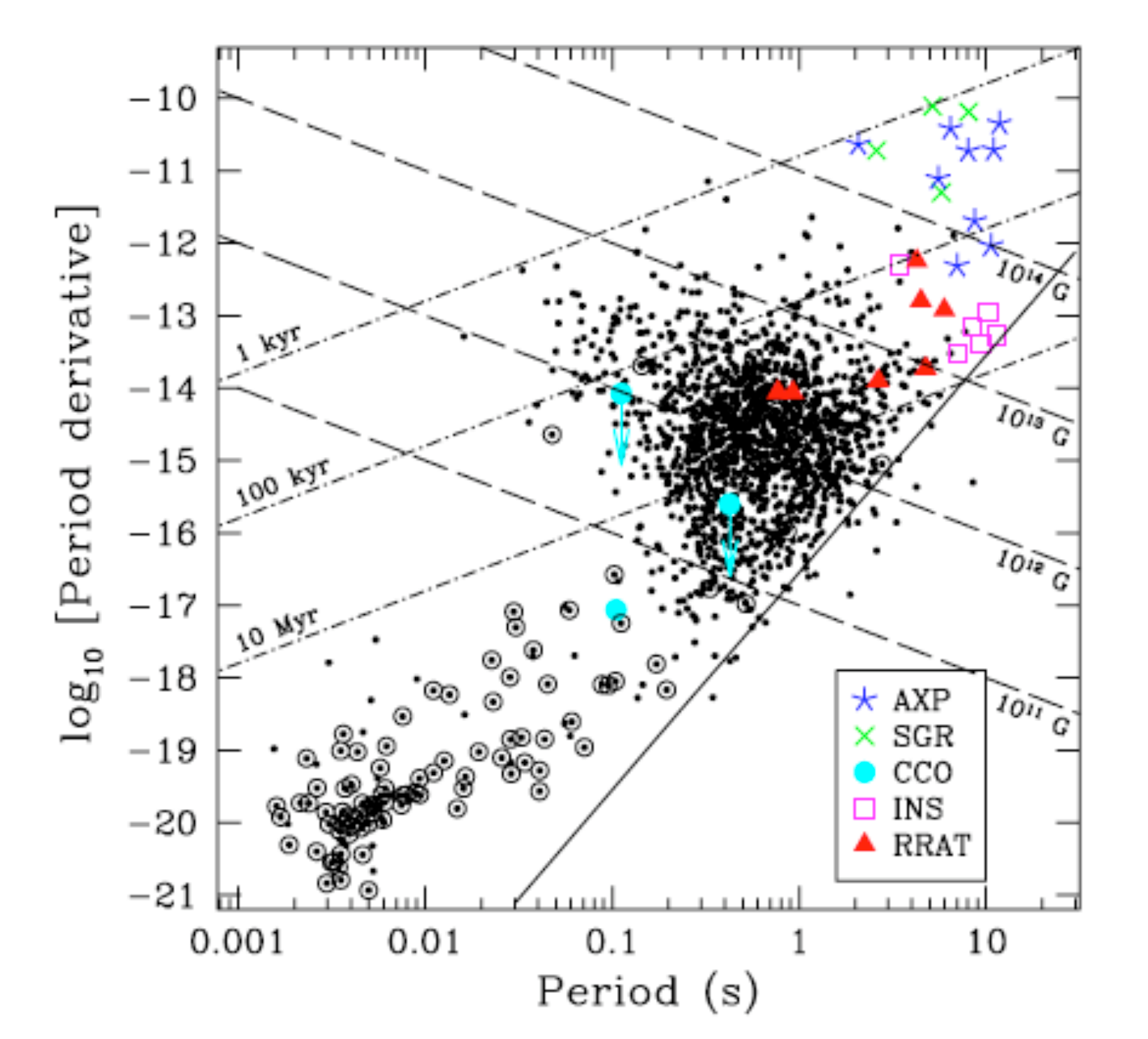}}}
\end{picture}
\end{center}
\vspace{7.5cm}
\caption{Observed RPP periods and period derivatives, from \cite{kaspi10}.   The line bounding the pulsar population corresponds to the magnetospheric voltage $\Phi = 10^{12}$ V, clearly marks a boundary beyond which pulsar emission is unlikely.  Simple estimates of pair creation  suggests this source of plasma should occur only for voltage larger than this value, which underpins the idea that pair creation is esential for radio emission [\cite{sturrock71}]. The change in slope of this ``death line'' al short period, small $\dot{P}$ indicates more sophistication in the pair creation physics, and/or in the association of pairs with radio emission, than is incorporated in the simplest models, a conclusion also apparent from the variety of quantitative problems with this widely accepted hypothesis \citep{hibsch01, medin10}.}
 \label{fig:PPdot}
\end{figure}
\noindent clear that the particle outflows are many orders of magnitude larger than the Goldreich-Julian current, thus motivating MHD models, where ${\boldsymbol E \cdot B} = 0$ is assumed from the start.  Pair creation occurring somewhere within the magnetospheres (first suggested by \citep{sturrock71}) is thought to be the origin of the required dense plasma, although a quantitative model that yields the high mass loss rates observed is still lacking [\citet{bucc10} and references therein].  In recent years, solutions for the MHD structure of the magnetosphere in the appropriate force-free limit have been obtained (numerically - the attempts of analytically minded theorists to guess the answer to the rather formidable free boundary/eigenfunction problem posed by the steady state version of the model,  uniformly failed over 30 years of attempts), first for the aligned rotator [\cite{contop99, gruzinov05, komiss06, timokhin06}], then for the full 3D oblique rotator [\cite{spit06, kala09}].

\begin{figure}[H]
\begin{center}
%\unitlength = 0.0011\textwidth
\hspace{10\unitlength}
\begin{picture}(300,100)(0,15)
\put(0,-100){\makebox(300,200)[tl]{\includegraphics[width=4in]{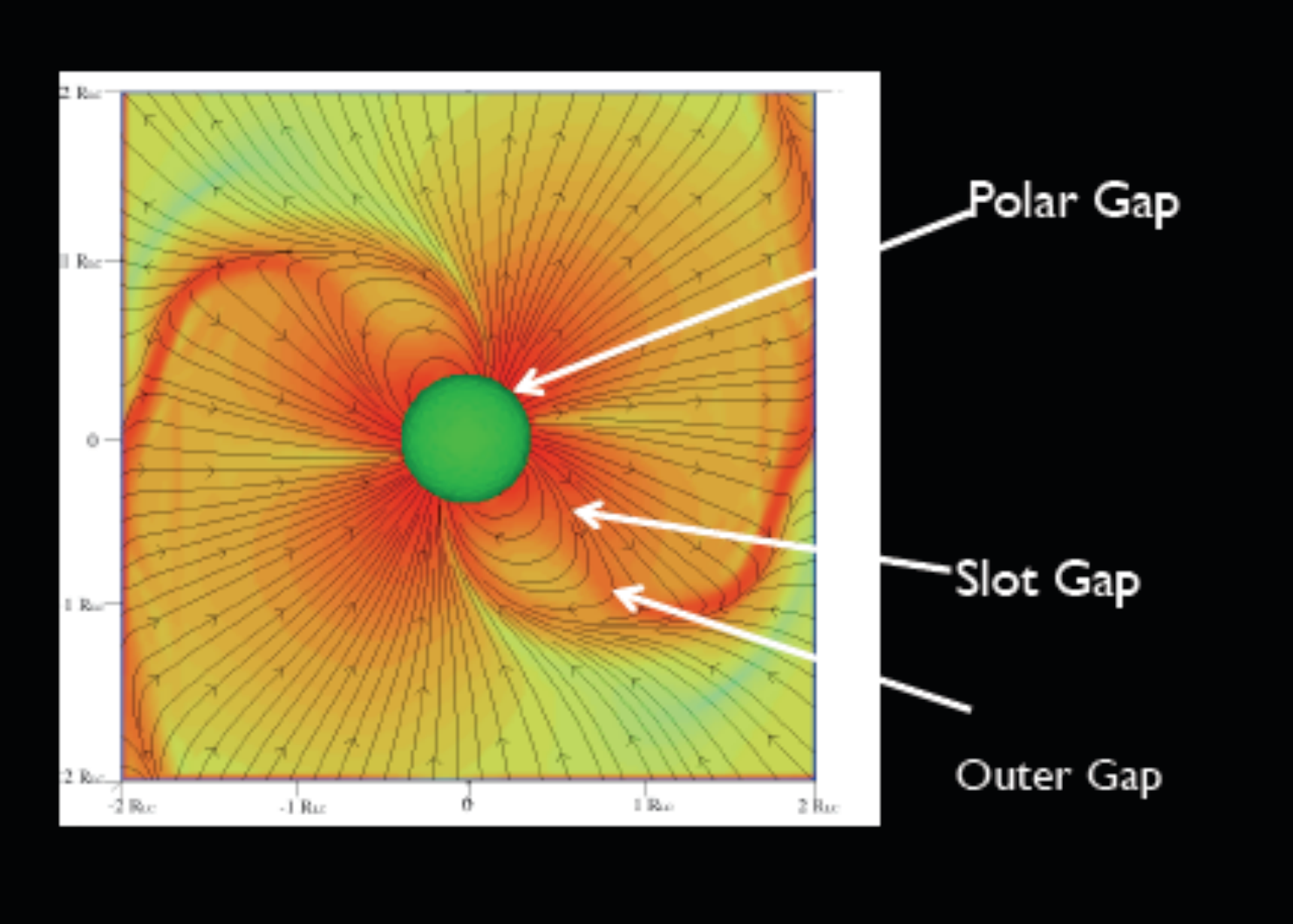}}}
\end{picture}
\end{center}
\vspace{5.8cm}
\caption{Field lines and current density of the oblique force-free rotator, $i = 60^\circ$, from Spitkovsky (2006), seen in a cut in the ${\boldsymbol{\Omega}, \boldsymbol{\mu}}$ plane. The last closed field lines end at a Y-point (a Y-line, in 3D) at distance $R_Y$ from the neutron star. The current sheet encloses and separates the closed from the open field regions of the magnetosphere, and the separate branches merge in the wind zone, where the folded sheet continues to separate the oppositely directed fields of the striped wind. The arrows and ``Gap'' labels locate sites where vacuum gaps have been postulated, in test particle  models of accelerators that lead to gamma ray emission and pair creation. None of these gaps appear in a current sheet accelerator based on the force-free magnetosphere model.}
 \label{fig:currentsheet}
\end{figure}

As far as the $P, \; \dot{P}$ diagram goes, the main result from these investigations is the innocuous looking conclusion in expression (\ref{eq:spindownnorm}). 
\begin{equation}
K = k(1 + \sin^2 i) \frac{\Omega^3 \mu^2}{c^3}, \;  k = 1 \pm 0.1.
\label{eq:spindownnorm}
\end{equation}
Physically, the most important result has been the identification of the current sheets separating the closed from the open regions of the magnetosphere, extending into the wind beyond the magnetosphere, whose last closed flux surface ends just touching the light cylinder whose cylindrical radius is $\varpi = c/\Omega $.  That such a current sheet should be present has been suspected from the early days of RPP research [{\it e.g.,} \cite{michel75}]. The error in $k$ reflects the uncertainties in the numerical treatment of the problem, many of which are associated with how the current sheet is represented in the numerical schemes. Figure \ref{fig:currentsheet} shows a slice through the  3d force-free magnetosphere of the $60^\circ$ rotator, from Spitkovsky's (2006) results. Observationally, expression (\ref{eq:spindownnorm}) shows that the vacuum rotator's ``braking index'' $n = 3$ is preserved in full force-free MHD, which contradicts the observed values in the small number of stars where $n$ has been determined \citep{livingstone07}, a contradiction which has led to a variety of suggestions ranging from evolution of the magnetic moment $\mu$ or the obliquity $i$ \citep{blandford88} to effects of reconnection on the rate of conversion of open magnetic flux to closed \citep{contop06}, as the star spins down and the closed zone expands at the expense of the amount of open magnetic flux. That reconnection might affect the braking index is readily derived from the fact that the torque really depends on the magnitude of the open magnetic flux.The amount of open flux depends on the size of the closed zone, which ends at $R_Y$.  If $R_Y/R_L < 1$, the torque increases because of more open field lines and larger Poynting  flux than is the case for a magnetosphere closing at $r = R_L$. \cite{bucc06} show that the braking index is 
\begin{equation}
n \equiv \frac{\Omega \ddot{\Omega}}{\dot{\Omega}^2} = 
    3 + 2 \frac{\partial \ln \left(1 + \frac{R_L}{R_Y} \right ) } {\partial \ln \Omega};
\label{eq:braking_recc}
\end{equation}
thus, If $R_Y/R_L$ decreases with decreasing $\Omega$, then $n<3$.  Reconnection usually is unsteady - figure \ref{fig:plasmoids} shows the blobs (``plasmoids'') ejected from the Y-line at the  \vspace*{-1cm}
\noindent base of the current sheet in the relativistic wind of the aligned rotator. The speculation is that spindown slightly biases these reconnection events so that on the much longer spindown timescale, the net open flux slowly converts to closed. These current fluctuations might be associated with the timing ``noise'' \citep{arons81a, cheng87} identified long ago with torque  fluctuations ({\it e.g.}, \citet{helfand80, scott03}), although recent analysis of longer data sets \citep{lyne10} in long period pulsars has called the noise interpretation into question.

In itself, the force-free model does not provide mechanisms for photon emission.  But it has a variety of implications, which are slowly being addressed.

\begin{itemize}
\item The model specifies the polar flux tube size and shape - it is noncircular with a polar cap center displaced from the magnetic axis, even when the magnetic field is the simplest, that of a star centered, point dipole \citep{bai10}. This has consequences for radio polarization structure, and for polar cap areas and dipole offsets inferred from soft X-ray emission from polar caps \citep{bogdanov07}, thought to be heated by magnetospheric particle bombardment \citep{arons81b, zavlin98, harding02}. These theoretical improvements of the polar cap model have yet to be noticed and incorporated in phenomenological models of the observations used by data analysts; such incorporation might yield interesting tests of the force-free model.
\vspace*{-2.75cm}
\begin{figure}[H]
\begin{center}
%\unitlength = 0.0011\textwidth
\hspace{10\unitlength}
\begin{picture}(300,110)(0,15)
\put(0,0.00){\makebox(300,200)[tl]{\includegraphics[width=4in]{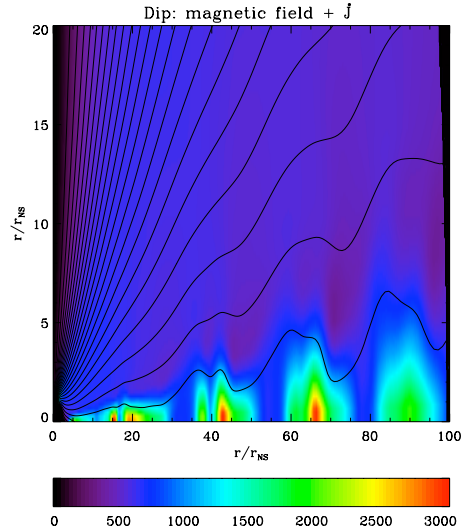}}}
\end{picture}
\end{center}
\vspace{5.8cm}
\caption{Plasmoids formed at the base of the current sheet of the relativistic aligned rotator, where the sheet crosses the light cylinder, from a relativistic MHD simulation of a newly born neutron star's magnetosphere \citep{bucc06}.  The dissipation that allows reconnection to occur is numerical. They move out radially at the local Alfven speed $v_A \approx c$, and recur on the magnetospheric Alfven transit time $\sim P /\pi$}
 \label{fig:plasmoids}
\end{figure}
\item The force free model quantitatively specifies the return currents required to prevent the star from charging up, as the polar flow extracts charge from the star. The results for the oblique rotator are partly in accord with long held expectations, that return current exists in a thin (``auroral'') sheet bounding the polar flux tube ({\it e.g.} \citep{gold69, michel75}), consistent with the open circuited model \citep{gold69} - current closure occurs far away, in the nebula/interstellar medium beyond the wind termination shock or perhaps in the outer wind, plus qualitatively new features: a) part of the return current surrounding the polar flux tube is spatially distributed, even in the aligned rotator; b) in the oblique rotator, this part of the current system is not a return current at all, but couple the two polar regions together \citep{bai10} - in the orthogonal rotator ($i = 90^\circ$), the auroral component of the current is entirely in the polar coupling flow, with the volume current out of each half of the polar cap having equal amount and opposite sign, also consistent with early expectations ({\it e.g.} \citet{saf78}) - the orthogonal rotator automatically balances it's charge loss. The radiative consequences of these features are as yet unexplored - for example, the spatially distributed part of he return current might be a good candidate for the site of ``conal'' component of pulsar radio emission, an idea which requires non-force modeling of the current flow and identification of a workable emission process within that current flow model\footnote{Some earlier ideas on this subject relating to field aligned acceleration and gamma ray emission can be found in \citep{arons83b, gruzinov08}, for example.} before one can relate the theoretical force free current distributions to the observations in a testable manner (although easier kinematic comparisons are certainly possible).
\item The location of the return current layer having been determined, the hypothesis that the return current layer is the site of the beamed particle accelerator that gives rise to the pulsed gamma rays observed by the FERMI and earlier orbiting gamma ray telescopes (see \cite{ray10} in these proceedings for a recent review) can now be investigated in the context of a self consistent magnetopspheric structure that allows a quantitative evaluation of the beaming characteristics implied by the radiating current sheet concept - see \citep{bai10} for a kinematical study of the radiating current sheet idea.

\end{itemize} 

Making progress on a physical model for radiation from the current layers can most expeditiously take advantage of the facts that a) pulsed gamma ray emission, when observed, is the largest photon output from rotation powered pulsars, but b)  generally has less luminosity than the spin-down luminosity of these stars. Figure \ref{fig:GammaEffic} illustrates this fact, which summarizes the results from the LAT instrument as of Spring 2010:
\vspace*{-2.5cm}
\begin{figure}[H]
\begin{center}
%\unitlength = 0.0011\textwidth
\hspace{10\unitlength}
\begin{picture}(300,110)(0,15)
\put(0,-150){\makebox(300,200)[tl]{\includegraphics[width=4in]{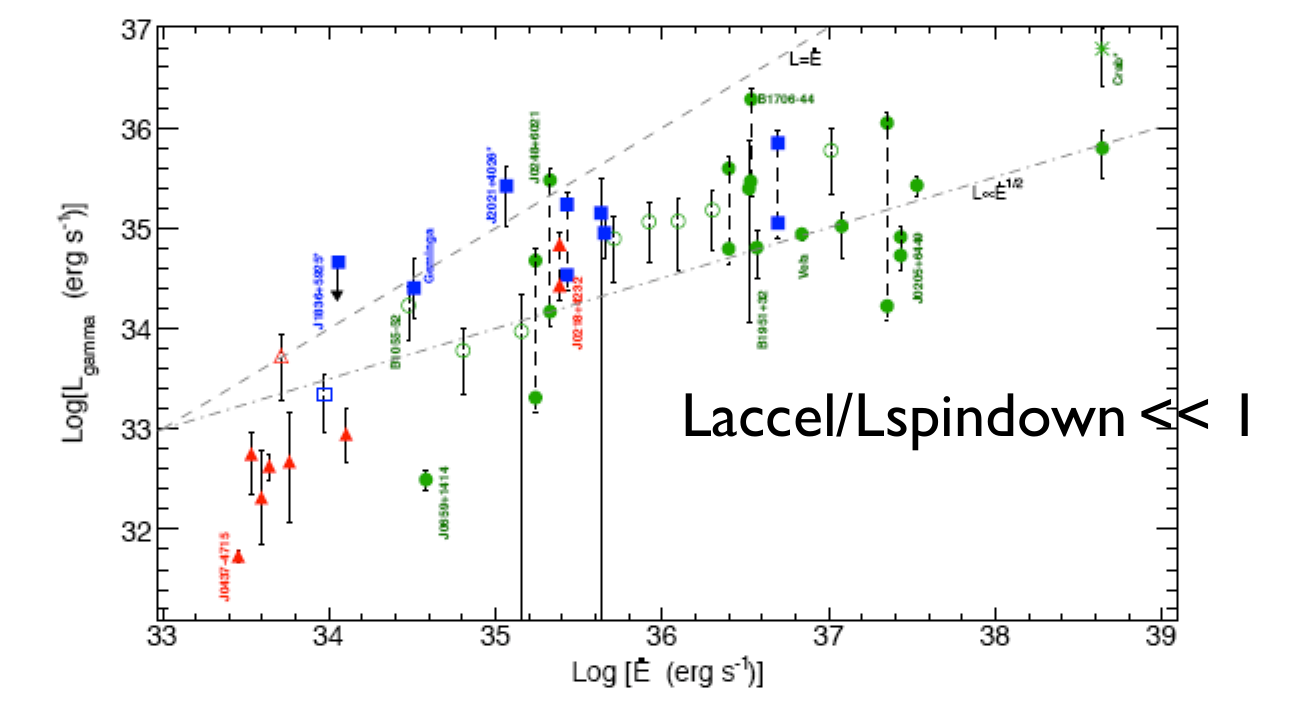}}}
\end{picture}
\end{center}
\vspace{4cm}
\caption{Ratio of observed gamma ray luminosity $L_{\rm gamma}$ to observed spindown power $\dot{E} = L_{\rm spindown}$ for LAT pulsars \citep{abdo10a}.  If the acceleration of the gamma ray emitting particles is radiation reaction limited (generally true in models that assume curvature radiation as the emission mechanism, as is the case in most of the ``gap'' models).  The trend at higher luminosities roughly follows the efficiency $L_{\rm gamma}/L_{\rm spindown} \propto (\sqrt{\dot{E}}^{-1}$ expected if some mechanism limits the accelerating voltage $\Delta \Phi$ to a fixed fraction of the magnetospheric voltage $\Phi_{\rm mag} = \sqrt{\dot{E}/c}$, [{\it e.g.}, \cite{arons96, harding02}], a limitation plausibly attributed to pair creation.  The large dispersion in Figure \ref{fig:GammaEffic} comes primarily from the uncertain distances, and also to the uncertain beaming correction required to go from the fraction of the sky illuminated by the beam to the total emission that would be detected by an observer covering $4\pi$ ster - the LAT team assumed uniform phase-averaged  beaming across the sky (1 sterradian).}
 \label{fig:GammaEffic}
\end{figure}
Thus the energy invested in particle acceleration is a small fraction of the enegy stored in Poynting fluxes, for radiation reaction limited acceleration, so that the force free model can be considered as a good zeroth order magnetospheric description.

The voltage limitation prediction $L_{\rm gamma} \propto \Phi_{\rm mag} \propto \sqrt{\dot{E}}$ is the same as $L_{gamma} \propto$ ``particle current'' = particle flux = Goldreich-Julian flux in unidirectional beam models of the polar electric current flow \citep{harding81} only if the  accelerator carries a fixed fraction of the total electric current that enters into the spin-down torque, independent of $P, \dot{P}$.  In the traditional slot or outer gap models, pair creation establishes the limitation of the accelerator to a thin sheet either in the outer magnetosphere \citep{ruderman86} or back in the polar cap \citep{arons83a} - in such models, the gap width $w_*$ of a model which successfully reproduces the sharply peaked light curves, projected onto the neutron star following the field the poloidal field lines, is necessarily small compared to the polar cap size, and varies with the pulsars' magnetic moment and spin parameters.  This is not an issue for a model based on the currents flowing in the return current layer, which necessarily carries the whole magnetospheric return current and include a substantial fraction in a thin current sheet, for most obliquities of the magnetic moment with respect to the rotation axis.

\begin{figure}[H]
\begin{center}
%\unitlength = 0.0011\textwidth
\hspace{10\unitlength}
\begin{picture}(300,110)(0,15)
\put(50,-150){\makebox(300,200)[tl]{\includegraphics[width=3in]{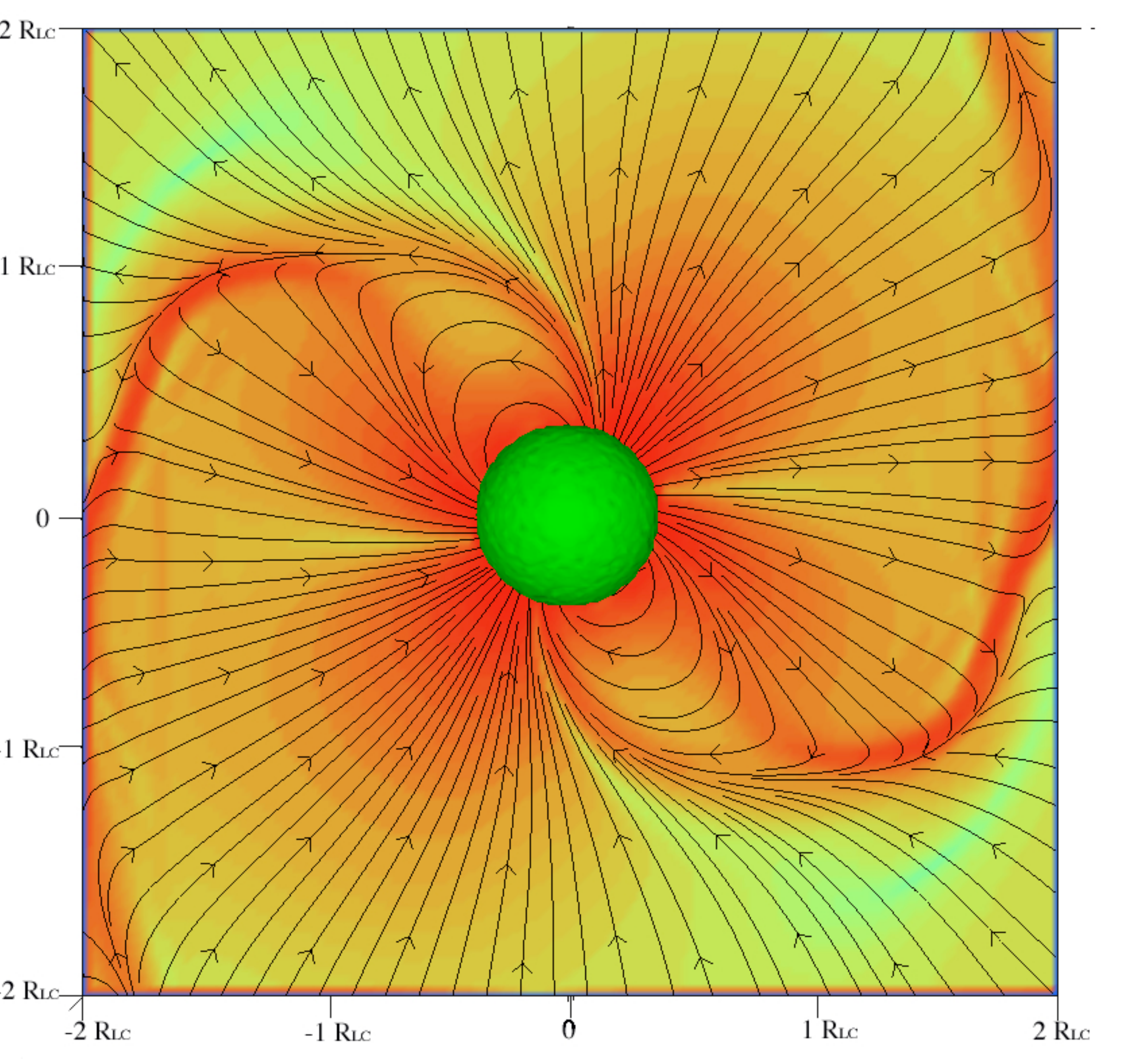}}}
\end{picture}
\begin{picture}(300,-110)(0,15)
\put(-20,-350){\makebox(300,200)[tl]{\includegraphics[width=4.65in]{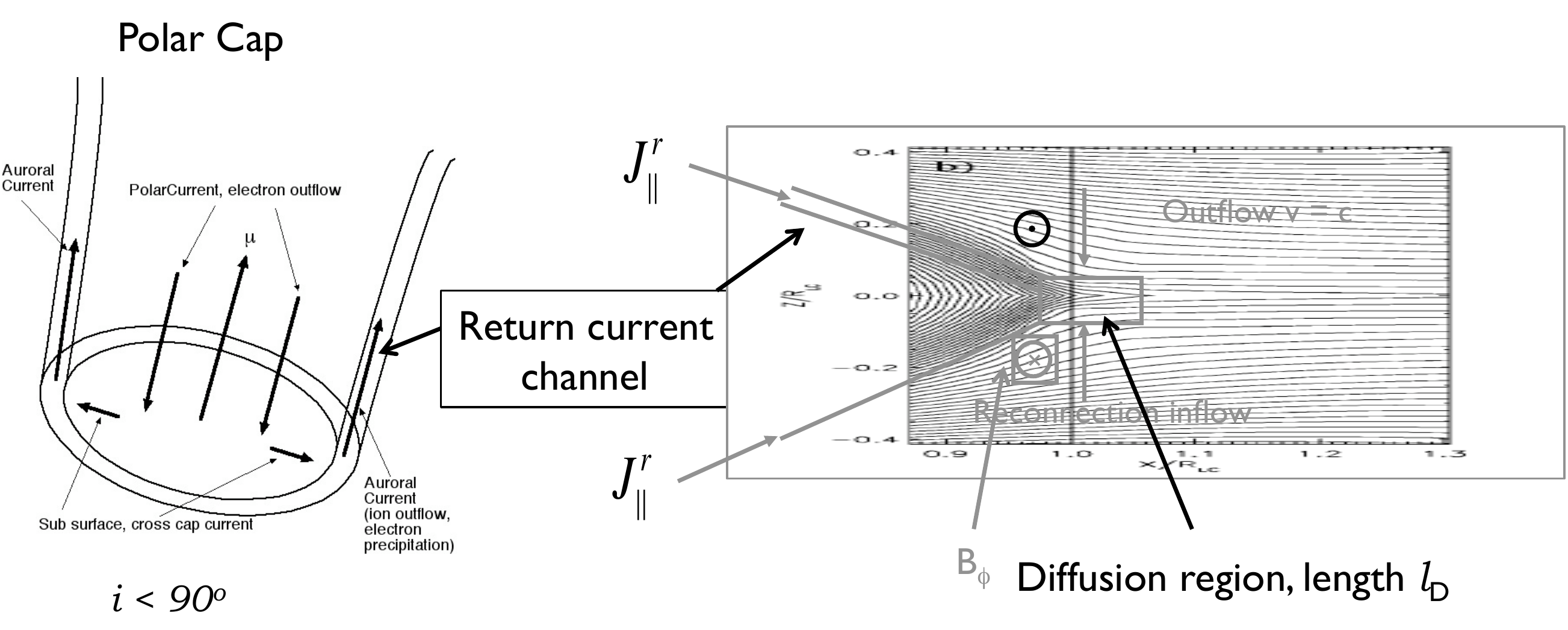}}}
\end{picture}
\end{center}
\vspace{10.5cm}
\caption{Upper Panel:  Electric current structure of the oblique force-free magnetosphere, inclination = $60^\circ$, from \cite{spit06}. The current sheet, indicated by the darker color, bounds the closed zone. The closed zone ends at a singular Y-line (in the ideal force-free electrodynamics approximation) at the light cylinder distance from the neutron star. Lower Panel, left: Current structure at the polar cap, illustrated for acute angle between the magnetic moment and the angular velocity (electron polar current, ${\boldsymbol \Omega} \cdot {\boldsymbol \mu} > 0$).  The return current in the current sheet consists of a precipitating electron beam, launched from the diffusion region around the Y-line and possibly augmented by high altitude pair creation within the current sheet, with the charges in the precipitating beam extracted by reconnection flow from the pair plasma flowing from the polar cap into the wind, plus a counterstreaming ion beam extracted electrostatically from the stellar atmosphere. In obtuse geometry (ion polar current, ${\boldsymbol \Omega} \cdot {\boldsymbol \mu} < 0$), positrons precipitate from the Y-line and counterstreaming electrons are extracted from the atmosphere. For clarity, the part of the return current not contained in the current sheet is omitted, even though this part of the current system is of increasing significance as $i \rightarrow 90^\circ$. Lower Panel, right: Possible structure of the Y-line region, with the termination of the closed zone to the left and the merger of the winds from the opposite polar caps to the right.  The ``guide field'' $B_\phi$ also reverses across the mid-plane of the flow, along with the poloidal open field.  As reconnection occurs, some of the ouflowing plasma (speed $c\beta, \; \beta \approx 1$) deflects toward the singular, unmagnetized ``diffusion'' region around the Y-line with speed $v_{rec} \sim 0.1 v_{\rm Alfven} = 0.1 c$. The figure represents a steady (in the co-rotating frame) flow model - in reality, the reconnection is likely to be bursty, as in Figure \ref{fig:plasmoids}, with formation of sporadic X-lines.}
 \label{fig:current_system}
\end{figure}

That magnetospheric current sheets, with particle densities within the sheets high compared to the Goldreich-Julian value, can sustain large parallel electric fields is well known in planetary magnetopsheres - such sheets are the accelerator sites of the particle beams that stimulate the aurora observed in the upper atmospheres of the Earth, Jupiter and Saturn, for example\footnote{For a review of such phenomena, see \cite{pasch02}. Of particular significance to pulsars is the fact that the field aligned currents that power narrow auroral arcs consist of precipitating electron beams launched from the reconnection region in the distant magnetotail and counterstreaming ions launched from the planetary atmosphere.}  An elementary illustration of this possibility comes from considering the inertial term in the generalized Ohm's for the electric field parallel to $\boldsymbol{B}$, which can be written as, in the relativistic case, as
\begin{equation}
E_\parallel = \frac{4\pi}{\omega_p^2}
     \left\{ 
      \frac{\partial J_\parallel}{\partial t} +
     \frac{{\boldsymbol {\vec B}}}{B} \cdot {\boldsymbol {\vec \nabla}} \cdot 
        \left[ \gamma 
        \left( {\boldsymbol {\vec J} {\vec v}} + {\boldsymbol {\vec v}{\vec J}} \right )
         \right] 
         \right\}
         \propto
         \frac{m\gamma}{n} \frac{I}{\Delta_{\rm current} \rho_B},
\label{eq:genOhm}
\end{equation}
with $m \gamma$ the particles' relativistic mass, $n$ their density, $I$ the total current set by the force free magnetosphere, $\Delta_{\rm current}$ is the thickness of the current carrying channel, and $\rho_B$ is the radius of curvature of the magnetic field. Thus inertia of the current carriers can act as a effective resistance in these high inductance systems, which establish the currents electromagnetically, forcing a parallel electric field to appear in the presence of any parallel (to B) load - pressure and radiation reaction itself are other effects  which can serve as loads, with pressure especially important in the diffusion region around the singular line shown in figure \ref{fig:current_system}. 
the fluid velocity.  Equation (\ref{eq:genOhm}) is most useful in the corotating frame, and when the current is due to relative motion between the species (electrons and positrons, and in some circumstances heavy ions) which is slow compared to the bulk fluid velocity {\bf v}.  In the current circumstance, it turns out that the current is better described as counter-streaming beams - in that case, describing the beams as separate fluids is more appropriate, and they can form the total plasma density in the current flow channel, rather than being a low density component in a much denser plasma.

Expression (\ref{eq:genOhm}) does make clear that acceleration is prone to maximize when the  the relativistic mass is high and the current density is high (large $I/\Delta_{\rm current})$. $\Delta_{\rm current}$ is generally microscopic, expected to be on the order of $c/\omega_p$, and is established by the dynamics of the the singular region where the closed zone ends, illustrated (in the cartoon approximation) in Figure \ref{fig:current_system}. The capture rate of pair plasma into the diffusion region is
\begin{equation}
\dot{N}^{in}_{{\rm diffusion}\pm} \approx \frac{2 l_D \Delta_L}{R_L^2} \frac{\beta_{\rm rec}}{\beta_{\rm wind}} \kappa_\pm \frac{c\Phi_{\rm mag}}{e},
\label{eq:capture}
\end{equation}
with $\kappa_\pm$ the multiplicity (multiplier of a fiducial Goldreich-Julian outflow rate $c\Phi_{\rm mag}/e$) that gives the number of {\it pairs} in the total outflow, $\beta_{\rm rec}$ the reconnection speed in units of $c$ and $\beta_{\rm wind}$ the polar wind outflow velocity (set equal to unity).  $\Delta_L$ is the thickness of the current channel at the light cylinder, assumed equal to the half height of the diffusion region, and $l_D$ is the length of the diffusion region. Pressure in the diffusion region expels the captured pairs from the diffusion region, outwards along the current sheet in the wind and inwards along the auroral current channels.  Expression (\ref{eq:capture}) assumes the plasma flux in the wind has a gradient across ${\boldsymbol B}$, as is likely since the accelerating electric field in the polar cap that leads to the pair creation that feeds the wind is small near the cap edge. The pressure supported electric field provides the accelerator which sorts the particles in the diffusion region into a precipitating beam (electrons in the geometry shown in Figure \ref{fig:current_system}) and an oppositely charge beam traveling outwards in the wind current sheet - the charge signs of the beams are as required by the global electrodynamics. The channel thickness is almost certainly comparable to the skin depth in the pairs. Taking the plasma gradient into account leads to the lower limit to the channel thickness
\begin{equation}
\Delta_{L{\rm min}} = \frac{c}{\omega_{p\pm} (\delta = c/\omega_{p\pm}(\delta)} \approx R_L \left(\frac{m_\pm c^2 \gamma_\pm}{2\kappa_\pm e \Phi_{\rm mag}} \right)^{1/3},
\label{eq:channel_width}
\end{equation}
where $\delta $ is the distance across B from the formal current sheet location, and a precipitating beam flux in the return current channel
\begin{equation}
F_\vee = \frac{l_D}{R_L} \kappa_\pm \frac{c\Phi_{\rm mag} /e}{2\pi R_L^2}\left(\frac{R_L}{r}\right)^3;
\label{eq:precip_flux}
\end{equation}
in the gradient case, $\Delta_L$ drops out of the precipitating flux evaluation. $\gamma_\pm$ is a measure of the four velocity and of he comparable four-velocity dispersion of the polar plasma flow emerging from the inner magnetosphere, predicted by pair cascade models to be on the order of $10^2$. Numerically, (\ref{eq:channel_width}) yields a very small value, on the order of meters to hundreds of meters, the specific value depending on $\Phi_{\rm mag}$ and $\kappa_\pm$.

The kinetics of relativistic reconnection being a largely untrodden subject, $l_D$ is more or less unknown - it could be as small as $\Delta_L$ itself (Petscheck style reconnection),  or as large as appears in the numerical dissipation  driven reconnection  observed in the force free  simulations, $l_D \sim 0.1 R_L$ (A. Spitkovsky, personal communication). In the terrestrial magnetosphere,  satellite observations suggest the diffusion region length is intermediate between the ion skin depth and the macroscopic scales.  For the discussion here (and in the more detailed report in preparation), treating $l_D$ as a parameter to be constrained by model comparisons to observations appears to be the wisest strategy. In analogy to observed non-elativistic reconnection, one expects $\beta_{\rm rec} \sim 0.1 v_{\rm Alfven}/c = 0.1 $, a value supported by the few PIC simulations of relativistic reconnection [{\it e.g.}, \cite{zenitani07}].

The precipitating particles form the ``hanging charge clouds'' invoked by \cite{gold69} to cause the electrostatic extraction of return current from the star's atmosphere, which happens if $n_\vee (R_*)  = F_\vee (R_*)/c \gg n_{GJ}(R_*)$, or , from (\ref{eq:precip_flux}),

\newpage
\vspace*{-2.8cm}
\noindent if 
\begin{equation}
\frac{l_D}{R_L} \frac{\beta_{\rm rec}}{\beta_{\rm wind}} \kappa_\pm \gg 1.
\label{eq:cloud}
\end{equation}
The condition in (\ref{eq:cloud}) is satisfied if $l_D$ is large compared to the skin depth, but still small compared to the numerical dissipation determined length observed in the force free simulations. It can be shown that  the consequent minimum total potential drop in the twin beam channel is 
\begin{eqnarray}
\Delta \Phi_{\rm min} & \approx & -\frac{1}{8} \Phi_{\rm mag} \frac{R_*}{R_L} 
                       \left( \frac{m_\pm c^2 \gamma_\pm}{2e \Phi_{\rm mag} \kappa_\pm} 
                        \frac{\beta_{\rm wind}}{\beta_{\rm rec}} \right)^{1/3} 
                          \frac{l_D}{\Delta_L} \cos i 
                        \nonumber \\                       
                       & =& -\frac{1}{8} \Phi_{\rm mag} \frac{R_*}{R_L} 
                            \left(\frac{\beta_{\rm wind}}{\beta_{\rm rec}} \right)^{1/3}
			 \frac{l_D}{R_L}  \cos i .
\label{eq:channel_potential}
\end{eqnarray}
If $l_D/R_L$ really is as large as 0.1, rather than comparable to the minimum scale $\Delta_L$, the accelerating potential in the return current channel is more than large enough (around $10^{13}$ Volts, in the Crab and Vela pulsars) to drive curvature gamma ray emission from the beams, which limits the particle energies by radiation reaction. Such energies are high enough to lead to pair creation, which may limit the acceleration, although as is clear from expression (\ref{eq:genOhm}), parallel potential drops in the current carrying region can be sustained even if the plasma is dense.  

If one models the gamma ray emission as being directly from the beams, one needs to take account of the electron beam going inwards, in acute geometry, where the upward beam is ions (generically, protons) - for obtuse geometry, the outbound beam is electrons, with an inbound positron beam.  Pair creation (through $\gamma - \gamma$ interactions with softer photons from the star), when it is important, makes the obtuse and acute geometries equivalent.

The radiation might be synchrotron emission, especially the nonthermal X-ray output.  The counterstreaming beams are electromagnetically two stream and shear unstable.  Since $\omega_{p,beam}$ can be comparable to the relativistic cyclotron frequency in the outer magnetosphere, the growing waves can excite finite Larmor gyration of the particles in the current channel, thus producing incoherent emission through hard X-rays - gamma rays are possible, under some circumstances. These X-rays are an alternative to soft photons from the star, as targets for $\gamma - \gamma$ pair production.  If the lower frequency waves can escape the plasma, they are a direct source of coherent emission, perhaps of interest to modeling giant radio pulses, which appear to come from the outer magnetosphere.

A more detailed description of the model will appear elsewhere.

\section{Follow the Mass}

Pulsar Wind Nebulae (PWNe) demonstrate that pulsars loss rest mass at a rate large compared to the fiducial electrodynamic particle loss rate $c\Phi_{\rm mag} /e$ [for a review, see \cite{gaensler06}]. The only known explanation is pair creation in the pulsars' magnetospheres. In many nebulae, the X-ray emitting particles rapidly lose energy to synchrotron radiation.  Then the nebulae are particle and energy calorimeters, allowing direct inference of the pair multiplicity in the wind, of TeV to PeV pairs.  The measured injection rates, up to $\sim 10^{38.5}$ pairs/s,  compare well to the predictions of existing pair creation models [{\it e.g.} \cite{hibsch01}], yielding multiplicities up to $\sim 10^4$. However, PWNe are also radio synchrotron emitters, radiation that samples much lower energy populations (100 MeV to 10 GeV), whose radiative efficiency is much less than their X-ray emitting cousins.  The result is a much larger population of pairs, whose radiative lifetime exceeds that of the nebulae.  The most efficient hypothesis is that these particles come from the embedded pulsars also, an idea supported by spectral continuity and by exotica, such as the observation of radio ``wisps'' near the Crab pulsars \citep{biet01}. Applying simple evolutionary models allows one to infer time averaged injection rates.  Recent evaluations by \cite{bucc10} and \citep{slane10} yield {\it lower limits} for multiplicities $\kappa_\pm$ all in excess of $10^5$  and {\it upper limits} for wind 4-velocities $\Gamma_{\rm wind} = \dot{E}_R /\dot{M}c^2 = e \Phi_{\rm mag}/2 \kappa_\pm m_\pm c^2$ all less than $10^5$ in a number of nebulae.  The data are the best for the younger systems, although even for these, the lack of far infrared data inhibits the analysis.\\
\newline
\begin{center}
\begin{tabular}{ccccc}
PWN Name & $\Phi_{\rm mag}$ (PV)  & Age (yr)  & $ \bar{\kappa}_\pm$  &  $\Gamma_{\rm wind}$ \\
 \hline
Crab &    100  & 955  &  $10^6$ & $5 \times 10^4$ \\
3C58 &  15  &  2100  & $ 10^{4.7}$ & $3 \times 10^4$ \\
B1509 & 120  & 1570 & $ 10^{5.3}$ & $1 \times 10^4$ \\
Kes 75 & 22 & 650 & $10^5$ & $7 \times 10^4$ \\
W44 & 1 & $20.3 \times 10^3$ & $ 10^5$ & $10^4 $ \\
K2/3 Kookaburra  & 5.5  & $13 \times 10^3$ & $10^5$ & $ 10^4$ \\
HESS J1640-465  &  3.5  & $  10^4 $ & $  10^6$ & $  10^{3.6}$
\end{tabular}
\end{center}

The inferred multiplicity excesses are a puzzle for pair creation theory [{\it e.g.} \cite{hibsch01, timokhin10}], perhaps resolvable by appealing to magnetic anomalies near the neutron stars' surfaces, the simplest being an offset of the dipole center from the stellar center, which strengthens the magnetic field at one pole \citep{arons98}. The increase this gives to the magnetic opacity can be greatly enhanced if the magnetic axis is also tipped with respect to the radial direction, since then gravitational bending of photon orbits with respect to the $B$ field direction much increases the magnetic opacity for pair creation. Such phenomenological modifications of the low altitude magnetic field must respect the observation that radio beaming morphology is consistent with the magnetic field being that of a star centered dipole quite close to the star \citep{rankin90, kramer98}. This problem warrants quantitative investigation.

The large inferred multiplicities imply the wind 4-velocities $\Gamma_{\rm wind}$ to be small compared to the much quoted value of $10^6$ inferred by \cite{kennel84} in their model of the Crab Nebula's optical and harder emission. The large mass loading and the inferred low wind four velocity has a large impact on the much storied ``$\sigma$ problem'' of pulsar winds.  In ideal MHD, the ratio $\sigma$ of magnetic energy to kinetic energy in the wind is conserved outside the fast magnetosonic radius (since for a cold flow the wind does not substantially accelerate outside this surface) and is large - even with the increased mass loading found from recent nebular studies, $\sigma$ is always well in excess of several hundred. Nevertheless the wind behaves at its termination shock as if $\sigma$ is small - MHD models of the nebulae suggest $\sigma $ at th termination shock is on the order of 0.02 in the Crab Nebula [{\it e.g} \cite{delzanna04}] and similar values are plausible in other systems. 

\begin{figure}[H]
\begin{center}
%\unitlength = 0.0011\textwidth
\hspace{10\unitlength}
\begin{picture}(300,-110)(0,15)
\put(-20,-230){\makebox(300,200)[tl]{\includegraphics[width=5in]{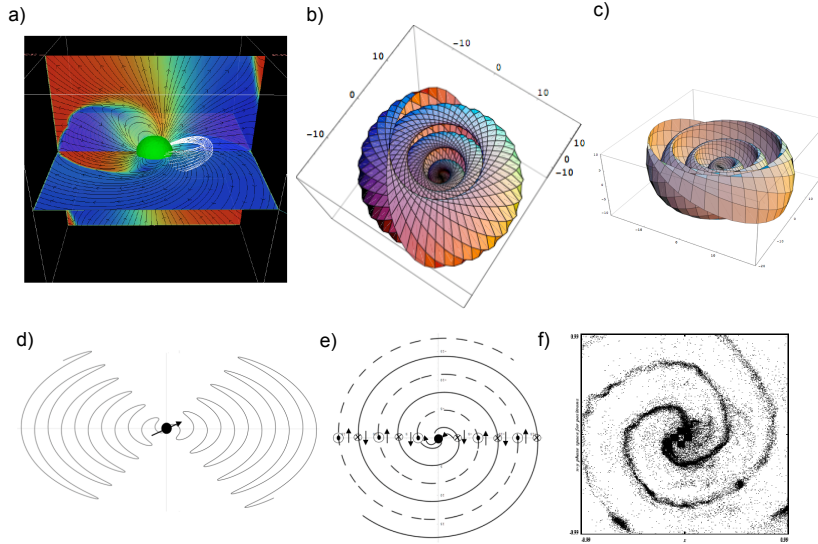}}}
\end{picture}
\end{center}
\vspace{10cm}
\caption{a) Magnetic Geometry of a Force-Free Rotator for $r<2R_L$, for $i = 60^\circ$,
from Spitkovsky (2006). The rapid transition to inclined split monopole field geometry for
$r > R_L$ is apparent. b) Geometry of the current sheet from the split monopole model for $i = 60^\circ$, $r > R_L$.  For clarity, only one of the two spirally wound current sheets is shown. As $i \rightarrow 90^\circ$, the sheets 
 almost completely enclose the star; for $r \gg R_L$, the spirals are tightly 
 wrapped ($B_r \ll B_\phi$) and the current sheet surfaces
 closely approximate nested spheres. 
  c) One sheet for $i=30^\circ$, shown for clarity. d) Meridional cross section
of the current sheet for $i=60^\circ$. e) Equatorial cross section snapshot of the current sheet, showing the two arm spiral form.  The arrows show the local directions of the 
magnetic field; the dots and crosses show the direction of the current flow. Panels b)-e) were constructed using Bogovalov's \citep{bogo99} analytic model of the asymptotic wind. f) Current sheet from a 2D PIC simulation of the inner wind, from unpublished work by Spitkovsky (used by permission),}
 \label{fig:stripes}
\end{figure}

\cite{coroniti90} suggested that because the wind of an oblique rotator has the magnetically striped structure shown in Figure \ref{fig:stripes}, magnetic dissipation of the corrugated B field, generically of a resistive nature propelled by instabilities of the current flow in the current sheet separating the stripes\footnote{ This wrinkled current sheet, frozen into the wind, is the second current sheet of this paper's title, but really it is the continuation of the sheet separating the closed and open zones interior to the light cylinder, as is apparent in Figure \ref{fig:current_system}.} might destroy the magnetic field of the wind interior to the termination shock, thus converting a high $\sigma$ flow into an effectively unmagnetized plasma.  If the current sheets separating the magnetic stripes are to merge with a speed $v_s < c$ as measured in the proper frame of the flow, before they reach the termination shock located at distance $R_{TS}$ from the neutron star, the merger time in the PWN frame $T_{\rm merge} = \pi \Gamma_{\rm wind}^2 (R_L/v_s)$ must be less than the flow time to the termination shock $R_{TS}/c$, therefore $\Gamma_{\rm wind} < \sqrt{(R_{TS}/\pi R_L) (v_s/c)} = 5 \times 10^4 \sqrt{v_s/c} \; ({\rm Crab})$ must be satisfied if Coroniti's model is to be viable.  This inequality is satisfied for multiplicities above $10^5$, which does appear to be the case for the young PWNe recently analyzed. In this low $\Gamma_{\rm wind}$, high pair multiplicity environment, complete destruction of the striped component of the wind's magnetic field is possible \citep{arons08}, contrary to the conclusion reached by \citep{lyub01}, but in accord with the conclusions of \citep{kirk03}, although for somewhat different reasons (drift kink instability dominate over tearing in pair plasmas).

Possible mechanisms that can lead to the necessary dissipation are drift-kink instability of the current sheet, considered as if it were a flat sheet \citep{zenitani07} and an interesting Weibel-like instability due to interaction between the sheets \citep{arons08}, an effect strongest in the equatorial sector where the folded sheet appears locally as neighboring flat sheets with antiparallel current flow in the latitude direction.  Most interesting from the astronomical perspective, the dissipation of the current sheets in the {\it inner} wind ($r \ll R_{TS}$, while responsible for only a small fraction of the magnetic destruction, may have sufficient luminosity to allow detection of VHE and UHE gamma-ray emission.  A much more elaborate paper describing these results is in preparation.
\vspace*{-2.8cm}

\noindent {\bf Acknowledgements}\\
\newline
The work described here has been supported by NSF grant AST-0507813, NASA grants NNG06GJI08G and NNX09AU05G and DOE grant DE-FC02-06ER41453. I have benefitted from discussions with E. Amato, N. Bucciantini, A. Spitkovsky and A.Timokhin.

\end{document}